\documentclass[preprint,aps,12pt,notitlepage,nofootinbib,tightenlines]{revtex4}
\usepackage{amsmath}
\usepackage{underscore}
\usepackage{bm}
\usepackage{times}
\usepackage{braket}
\usepackage{color}
\usepackage{epsfig}
\usepackage{slashed}
\usepackage{hyperref}
\usepackage{multirow}
\usepackage{booktabs}
\usepackage{array}
\usepackage{float}
\newcommand{\beq}{\begin{eqnarray}}
\newcommand{\eeq}{\end{eqnarray}}
\newcommand{\be}{\begin{equation}\begin{aligned}}
\newcommand{\ee}{\end{aligned}\end{equation}}

\newcommand{\gev}{\text{GeV}}

\definecolor{Red}{rgb}{1.,0.,0.}

\definecolor{Blue}{rgb}{0.,0.,1.}

\definecolor{nicered}{rgb}{0.7,0.1,0.1}
\definecolor{nicegreen}{rgb}{0.1,0.5,0.1}
\def\lsim{ {\ \lower-1.2pt\vbox{\hbox{\rlap{$<$}\lower6pt\vbox{\hbox{$\sim$}}}}\ } }
\def\gsim{ {\ \lower-1.2pt\vbox{\hbox{\rlap{$>$}\lower6pt\vbox{\hbox{$\sim$}}}}\ } }

\hypersetup{colorlinks,citecolor=nicegreen,linkcolor=nicered}
\begin{document}
%%%%%%%%%%%%%%%%%%%%%%%%%%%%%%%%%%%%%%%%%%%%%
\title{Pair production of the singlet vector-like B quark at the CLIC}
\author{Jin-Zhong Han$^{1}$\footnote{E-mail: hanjinzhong@zknu.edu.cn}, Yao-Bei Liu$^{2}$\footnote{E-mail: liuyaobei@hist.edu.cn}, Shi-Yu Xu$^{1}$}
\affiliation{1. School of Physics and
Telecommunications Engineering, Zhoukou Normal University, Zhoukou 466001, P.R. China\\
2. Henan Institute of Science and Technology, Xinxiang 453003, P.R. China}

\begin{abstract}
Vector-like quarks~(VLQs)  are a common feature of many scenarios of new physics  beyond the Standard Model~(SM), which generally decay into a SM third-generation quark with a SM gauge boson, or a Higgs boson. The presence of a new exotic decay mode of VLQs will reduce the branching ratios of these standard decay modes and thus relax the current mass exclusion limits from LHC experiments.
Based on a model-independent framework,
we investigate the prospect of discovering the pair production of the weak-singlet VLQ-$B$ at the future 3-TeV Compact Linear Collider~(CLIC), by focusing on the final states including one $Z$ boson and four $b$-jets via two types of modes: $Z\to \ell^{+}\ell^{-}$ and $Z\to \nu\bar{\nu}$.
 By performing a rapid detector simulation of the signal and background events, and considering the initial state radiation and beamstrahlung effects,
the  exclusion limit at  the 95\% confidence level  and the $5\sigma$ discovery prospects are respectively obtained on the branching ratio of $B\to bZ$ and the VLQ-$B$ masses at the future 3-TeV CLIC with an integrated luminosity of 5 ab$^{-1}$.
\end{abstract}

\maketitle
\newpage
\section{Introduction}
New heavy quarks appear in a variety of new physics models beyond the Standard Model~(SM),  formulated to obtain an answer to the problem of the naturalness of the electroweak~(EW) scale~\cite{DeSimone:2012fs,ArkaniHamed:2002qy,Han:2003wu,Chang:2003vs,Agashe:2004rs,He:1999vp,Wang:2013jwa,He:2001fz,He:2014ora}.
Vector-like quarks~(VLQs) are spin 1/2 particles characterized by having left- and right-handed components
defined by the same color and EW quantum numbers~\cite{Aguilar-Saavedra:2013qpa}, and thus  could still be viable under the present searches. However,  an extra fourth generation of SM-like quarks~\cite{He:2001tp,Chen:2012wz} should be much heavier due to the EW precision constraints, and thus a certain non-perturbative method is needed to reliably analyze the strongly coupled Yukawa sector of these extra heavy chiral quarks. VLQs can have different charge assignments under the SM EW
gauge group $SU(2)_{L}\times U(1)_{Y}$. Hence, there exists the possibility of having multiple VLQs, including electroweak singlet [$T$, $ B$], electroweak doublets [ $\left(X,T\right),\left(T,B\right)$ or $\left(B,Y\right)$], or electroweak triplets [$\left(X,T,B\right)$ or $\left(T,B,Y\right)$]. It is generally assumed that the VLQs decay into a SM third-generation
quark with a SM gauge boson or a Higgs boson, therefore providing a rich phenomenology at future high-energy colliders~\cite{Atre:2011ae,Cacciapaglia:2011fx,Nutter:2012an,Okada:2012gy,Buchkremer:2013bha,Yang:2014usa,Matsedonskyi:2014mna,Backovic:2014uma,Barducci:2017xtw,Cacciapaglia:2018lld,Cacciapaglia:2018qep,
Liu:2017sdg,Moretti:2017qby,Liu:2019jgp,
Buckley:2020wzk,Deandrea:2021vje,Tian:2021oey,Tian:2021nmj,Yang:2021btv,Han:2022npb,Han:2022jcp,Han:2023jzm,Alves:2023ufm}.

For a vector-like $B$-quark~(VLQ-$B$)  with electric charge -(1/3)e, direct searches generally assume three standard decay channels: $B\to tW$, $bZ$, and $bH$. Very recently, ATLAS and CMS
collaborations  primarily focus on the quantum chromodynamics~(QCD)-induced  pair production modes of VLQs and lead to lower bounds on the VLQ masses of approximately 1-1.5~TeV~\cite{ATLAS:2018tnt,ATLAS:2018ziw,CMS:2018zkf,CMS:2019eqb,CMS:2020ttz,ATLAS:2021ibc}.
Using Run 2 data with a total integrated luminosity of 137 fb$^{-1}$, the CMS Collaboration recently presented a search for VLQ-$B$ pair production in the fully hadronic final state ~\cite{CMS:2020ttz}, and excluded their masses up to 1.57  and 1.39~TeV for 100\% $B\to bh$ and 100\% $B\to tZ$, respectively.
  The ATLAS Collaboration presented a search for the pair production of VLQs optimized for decays into a $Z$ boson and a third-generation SM quark~\cite{ATLAS:2022hnn}. The lower limits on the masses of  VLQ-$B$ are 1.20~TeV for the weak-isospin singlet model and 1.42 TeV  for 100\%  $B\to tZ$ cases. However, these bounds would be relaxed
 if such VLQs were to have non-standard decay channels.
Recently, exotic decays of the VLQs in different set-ups with different collider
signatures have been considered in the literature~\cite{Aguilar-Saavedra:2017giu,Das:2018gcr,Benbrik:2019zdp,Cacciapaglia:2019zmj,Aguilar-Saavedra:2019ghg,Zhou:2020byj,Wang:2020ips,Corcella:2021mdl,Cacciapaglia:2021uqh,Cui:2022hjg,Banerjee:2022izw,Banerjee:2022xmu,Bhardwaj:2022wfz,Bhardwaj:2022nko,Bardhan:2022sif}.

Compared with the complicated QCD background at the
hadron colliders, the future high-energy linear $e^{+}e^{-}$ collider is a precision machine with which the properties
of such new VLQs can be measured precisely~\cite{Kong:2007uu,Senol:2011nm,Guo:2014piv,Liu:2014pts}. In particular, the final stage of Compact Linear Collider~(CLIC) will operate at  energy of 3 TeV~\cite{CLICDetector:2013tfe}, and any such new particles can be produced with sizable rate up to the
kinematic limit of 1.5 TeV, and in some cases up to 3 TeV, via single production
mechanisms~\cite{Franceschini:2019zsg}.
Recently, the single production processes of VLQs at the CLIC have been widely studied via different decay modes~\cite{Qin:2021cxl,Han:2021kcr,Han:2021lpg,Qin:2022mru,Han:2022exz,Han:2022zgw,Han:2022rxy,Yang:2023wnv}.
Unlike the single production mode, the production of VLQs pairs is model-independent, i.e.,  their cross sections depend only on their masses~\cite{Qin:2023zoi}.
In this paper we will focus on the illustrative examples of VLQ-$B$ pairs interacting with the third generation of SM quarks, and we will analyze their pair production signatures via the  standard decay channels $B\to bZ~(bh)$ at the future 3-TeV CLIC.  Furthermore, we will estimate the reach for
discovering (or excluding) VLQ-$B$ in which it is assumed that the exotic decay mode is possible and take the branching ratio~(BR) of $B\to bZ$ as a free parameter.

This paper is organized as follows: in Sec. II, we briefly describe  the couplings of the singlet VLQ-$B$ with the SM particles  in the simplified model and the direct
LHC constraints on its mass and the branching ratio of $B\to bZ$. In section III,  we discuss the pair production process at the 3-TeV CLIC, and perform a detailed collider analysis
of the relevant signals and backgrounds. Finally, we provide a summary in Sec. IV.

\section{Singlet VLQ-$B$ in the simplified model}
A generic parameterization of an effective Lagrangian for the singlet VLQ-$B$ is given by~\footnote{
Details are provided on the URL \href{http://feynrules.irmp.ucl.ac.be/wiki/VLQ\_bsingletvl}{\texttt{http://feynrules.irmp.ucl.ac.be/wiki/VLQ\_bsingletvl}.}}
\beq
{\cal L}_{\rm eff} =&& \frac{g\kappa_B}{\sqrt{2}}[\frac{1}{\sqrt{2}}\bar{B}_{L}W_{\mu}^{+}
    \gamma^{\mu} t_{L}+
    \frac{1}{2c_W}\bar{B}_{L} Z_{\mu} \gamma^{\mu} b_{L}
    - \frac{m_{B}}{2m_{W}}\bar{B}_{R}hb_{L} -\frac{m_{b}}{2m_{W}} \bar{B}_{L}hb_{R} ]- \nonumber\\
    &&
    \frac{e}{6c_{W}}\{\bar{B}B_{\mu}\gamma^{\mu}B\}-
    \frac{e}{4s_{W}}\{\bar{B}W^{3}_{\mu}\gamma^{\mu}B\}+ h.c.,
  \label{BsingletVL}
\eeq
where $g$ is the $SU(2)_L$ gauge coupling constant, and there are two free parameters: the VLQ-$B$ quark mass $m_B$ and  the coupling strength to SM quarks in units of standard couplings, $\kappa_B$.

\begin{figure}[thb]
\begin{center}
\vspace{-0.5cm}
\centerline{\epsfxsize=10cm \epsffile{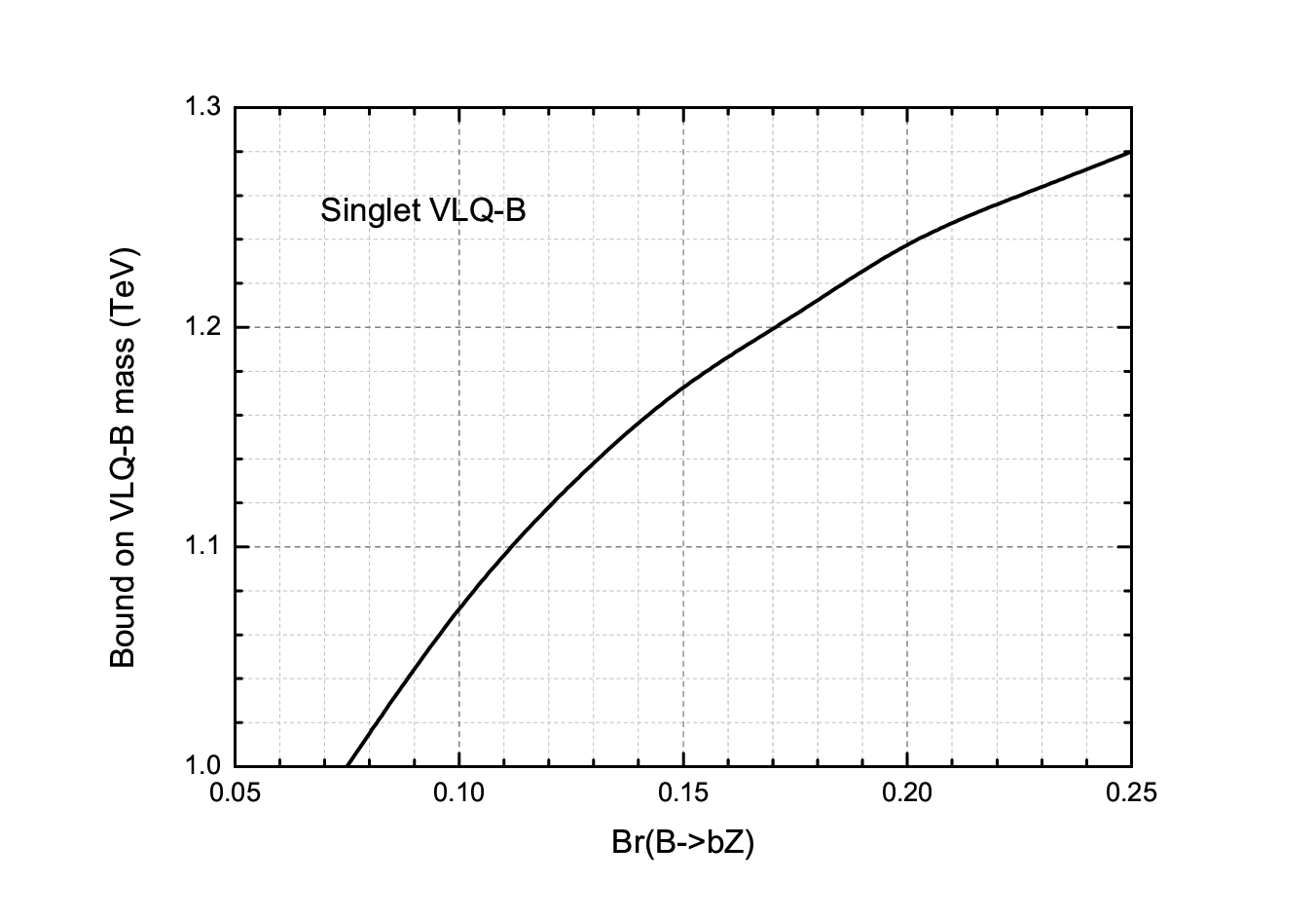}}
\caption{LHC exclusion limits on the VLQ-$B$ as a
function of $Br(B\to bZ)$  in the singlet $B$ model including non-standard decay modes.}
\label{bound-mass}
\end{center}
\end{figure}

Assuming an almost degenerate VLQ mass hierarchy, they are generally assumed to decay into a third-generation quark and either a $W/Z$ boson or a Higgs boson.
 For a heavy weak-isospin singlet VLQ-$B$, the relationship of the BRs of three standard decay modes is
  \beq \label{relationship}
   {\rm Br}(B\to tW)\approx {\rm 2Br}(B\to bZ)\approx {\rm 2Br}(B\to bh),
   \eeq
    which is a good approximation as expected from the Goldstone
boson equivalence theorem~\cite{He:1992nga,He:1993yd,He:1994br,He:1996rb,He:1996cm}.
 With the introduction of the new decay modes $B\to X$, the sum of  the above three  standard decay modes changes to
 \beq\label{beta}
{\rm Br}(B\to bZ)+{\rm Br}(B\to bh)+{\rm Br}(B\to tW) =1-\beta_{new},
 \eeq
where $\beta_{new}$ is the BR for the new exotic decay channel, such as $B\to bS$~(an additional scalar or pseudoscalar particle~\cite{Bhardwaj:2022nko,Bardhan:2022sif}).  A smaller BR in the new mode $\beta_{new}$ implies larger
BRs in the SM modes. Based on Eqs.~(\ref{relationship}-\ref{beta}),  we can obtain
\beq\label{brbz}
{\rm Br}(B\to bZ)\simeq( 1-\beta_{new})/4.
\eeq
In this case, it is instructive to analyze the current constraints from direct searches of VLQ-$B$.
Very recently, Refs.~\cite{Bhardwaj:2022nko,Bardhan:2022sif} recast the current LHC searches to put mass exclusion bounds on VLQ-$B$ as a function of the branching ratio in the new decay mode.
In Fig.~\ref{bound-mass}, we show the exclusion mass limits on VLQ-$B$ as a function of ${\rm Br}(B\to bZ)$ according to Eq.~(\ref{brbz}) and the results in the literature~\cite{Bhardwaj:2022nko}, where they recast the relevant limits from
the available exclusive~\cite{CMS:2019eqb, CMS:2020ttz} and inclusive~\cite{ATLAS:2021ibc} searches to select the strongest one. From the rescaled VLQ-$B$ limits in Fig~\ref{bound-mass}, one can see the VLQ-$B$ mass could be smaller than 1.3 TeV for the smaller BR of the standard decay mode $B\to bZ$, which implies that the VLQ-$B$ could be pair produced at the future 3-TeV CLIC.

%%Fig.1 %%%%%%%%%%%%%%%%%%%%
%%%%%%%%%%%%%%%%%%%%%%%%%
\section{Collider simulation and analysis}
\begin{figure}[thb]
\begin{center}
\vspace{-0.5cm}
\centerline{\epsfxsize=10cm \epsffile{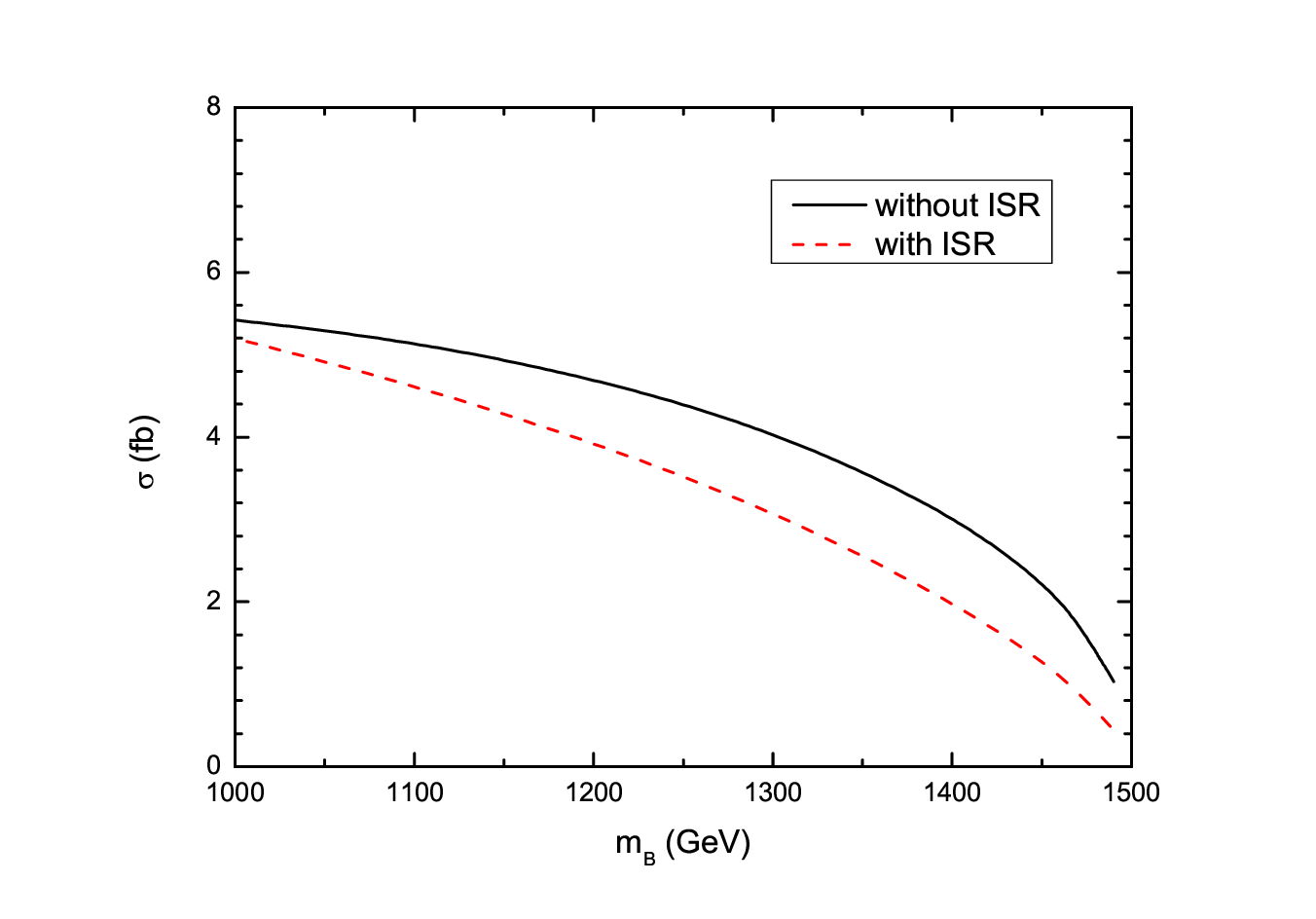}}
\caption{Total cross section of $\sigma$ as a function of $m_B$ at the 3-TeV CLIC with and without ISR effects.}
\label{cs}
\end{center}
\end{figure}

In order to make a prediction for the signal, we calculate the production cross section for the process $e^{+}e^{-}\to B\bar{B}$ at leading order~(LO).  Note that here the effects of  initial state radiation (ISR) and beamstrahlung are also considered at the 3-TeV CLIC  in MadGraph5\_aMC\_v3.3.2~\cite{mg5} by adding the following commands in the run\_card:
\begin{verbatim}
  set lpp1 +3
  set lpp2 -3
  set pdlabel clic3000ll
\end{verbatim}

In Fig.~\ref{cs}, we show the dependence of the cross sections $\sigma$  as a function of  $m_B$ with~(without)  ISR and beamstrahlung effects. One can see that the cross sections can be changed with ISR and beamstrahlung effects compared with those without ISR and beamstrahlung effects. In the region of $m_B\in$ [1000, 1490]~GeV, the ratios of cross sections with and without ISR and beamstrahlung effects are changed from 0.96 to 0.43;  thus, it is necessary to consider these effects at the future 3-TeV CLIC, especially for the high VLQ-$B$ masses. For $m_B=1.2~(1.45)$ TeV, the cross section can reach about 3.94~(1.29) fb with ISR and beamstrahlung effects.
%%Fig.2 %%%%%%%%%%%%%%%%%%%%
%%%%%%%%%%%%%%%%%%%%%%%%%

 Under the narrow-width approximation (NWA), it is possible
to separate and factorize production and decay of the heavy quarks, thus allowing for a model-independent analysis of the results~\cite{Moretti:2016gkr}. For the processes $e^{+}e^{-}\to B\bar{B}\to bZ\bar{b}Z$ and $e^{+}e^{-}\to B\bar{B}\to bZ\bar{b}h$,  their cross sections can be written as
\be
\sigma_{bZbZ}\equiv \sigma_{e^{+}e^{-}\to B\bar{B}}\times {\rm Br}(B\to bZ)\times {\rm Br}(B\to bZ),\nonumber\\
\sigma_{bZbh}\equiv \sigma_{e^{+}e^{-}\to B\bar{B}}\times {\rm Br}(B\to bZ)\times {\rm Br}(B\to bh).
\ee
Assuming the relationship of ${\rm Br}(B\to bZ)\simeq {\rm Br}(B\to bh)$, we will take ${\rm Br}(B\to bZ)$ as a free parameter in the remainder of this article.

We demand that a pair-produced VLQ-$B$ event should have at least
one $Z$ boson and four $b$-jets,  whith one pair of $b$-jets decaying from a $Z$ boson or a Higgs boson,
\begin{equation}
    e^{+}e^{-} \to  B\bar{B} \to \left\{ \begin{array}{l} \left(b Z\right)\,\left(\bar{b}Z\right) \\ \left(b Z\right)\,\left(\bar{b}h\right) \\ \left(b h\right)\,\left(\bar{b}Z\right) \end{array}\right. \to b\bar{b}b\bar{b}Z.
\end{equation}

In the next section, we will perform the Monte Carlo simulation and explore the discovery potentiality of VLQ-$B$
through the subsequent leptonic decay channel $Z\to \ell^{+}\ell^{-}$ and the invisible decay channel $Z\to \nu\bar{\nu}$, respectively.
To generate events for each signal benchmark, we pick
model parameters such that $\kappa_B=0.1$ and ${\rm Br}(B\to bZ)=0.25$ while ensuring
 that the NWA remains valid for VLQ-$B$.

Monte Carlo event simulations for the signal and SM background are interfaced to Pythia 8.20~\cite{pythia8}  for fragmentation and showering. All event samples are fed into the Delphes 3.4.2 program~\cite{deFavereau:2013fsa} with the CLIC detector card designed for 3 TeV~\cite{Leogrande:2019qbe}. In our analysis, jets are clustered with the Valencia Linear Collider~(VLC) algorithm~\cite{Boronat:2014hva,Boronat:2016tgd} in exclusive mode with a fixed number of jets~($N=4$ where $N$ corresponds to the number of partons expected in the final state) and fixed-size parameter $R=0.7$. The $b$-tagging efficiency is taken as the loose working points with 90\% b-tagging efficiency in order not to excessively reduce the signal efficiency.  Finally, event analysis is performed by using MadAnalysis 5~\cite{ma5}.

\subsection{The decay channel $Z\to \ell^{+}\ell^{-}$}
In this subsection, we analyze the signal and background events at the 3-TeV CLIC through the $Z\to \ell^{+}\ell^{-}$~($\ell=e,\mu$) decay channel:
\begin{equation}
    e^{+}e^{-} \to B\bar{B} \to \left\{ \begin{array}{l} \left(b Z\right)\,\left(\bar{b}Z\right) \\ \left(b Z\right)\,\left(\bar{b}h\right) \\ \left(b h\right)\,\left(\bar{b}Z\right) \end{array}\right. \to 4b+\ell^{+}\ell^{-}.
\end{equation}
For this channel, the typical signal is two opposite-sign same-flavor~(OSSF) leptons coming from one $Z$ boson and four $b$-tagged jets, with a pair of $b$-tagged jets coming from a $Z$ boson or a Higgs boson. The dominant SM backgrounds come from the SM processes  $e^{+}e^{-}\to Zhb\bar{b}$ and $e^{+}e^{-}\to ZZb\bar{b}$. Note that the contributions from the processes $e^{+}e^{-}\to ZZh$, $e^{+}e^{-}\to ZZZ$, and $e^{+}e^{-}\to Zhh$ are also included with the decay modes $Z\to \ell^{+}\ell^{-}$, $Z\to b\bar{b}$, and $h\to b\bar{b}$.

To identify objects, we choose the basic cuts at parton level for the signals and SM backgrounds as follows:
 \be
p_{T}^{\ell/j/b}>~25~\gev, \quad
 |\eta_{\ell/b/j}|<~2.5, \quad
 \Delta R_{ij} > 0.4, \\
  \ee
where  $\Delta R=\sqrt{\Delta\Phi^{2}+\Delta\eta^{2}}$ denotes the separation in the rapidity-azimuth plane, and $p_{T}^{\ell, b, j}$ are the transverse momentum of leptons, $b$-jets, and light jets.
 %%%%%%%%%%%%%%%%%%%%%%%%%%%%%%%%%%%%%%%%%%%%%%%%%%
\begin{figure*}[htb]
\begin{center}
\centerline{\hspace{2.0cm}\epsfxsize=9cm\epsffile{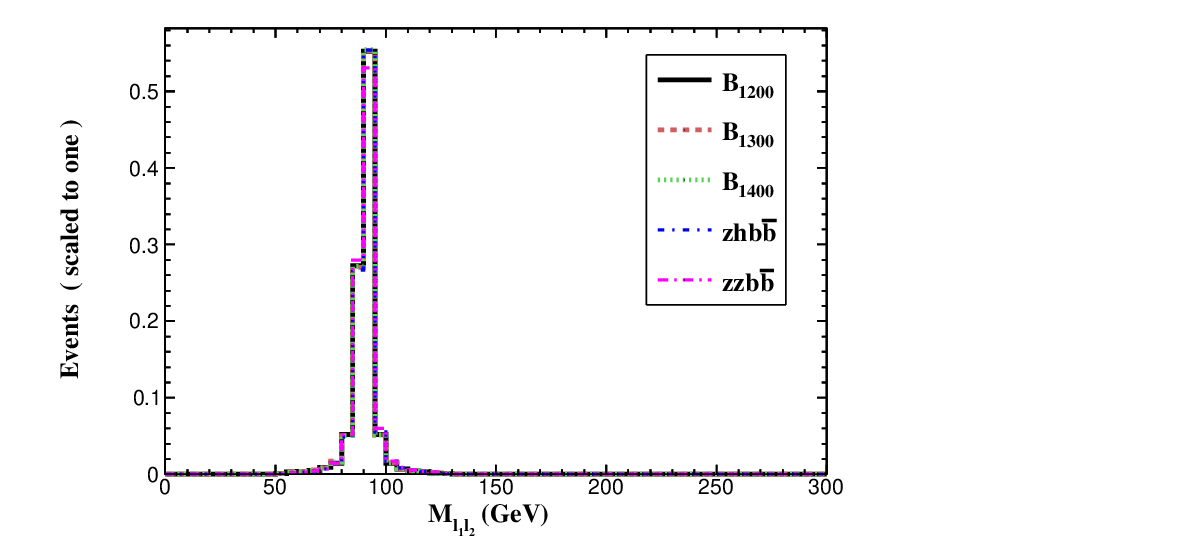}
\hspace{-2.0cm}\epsfxsize=9cm\epsffile{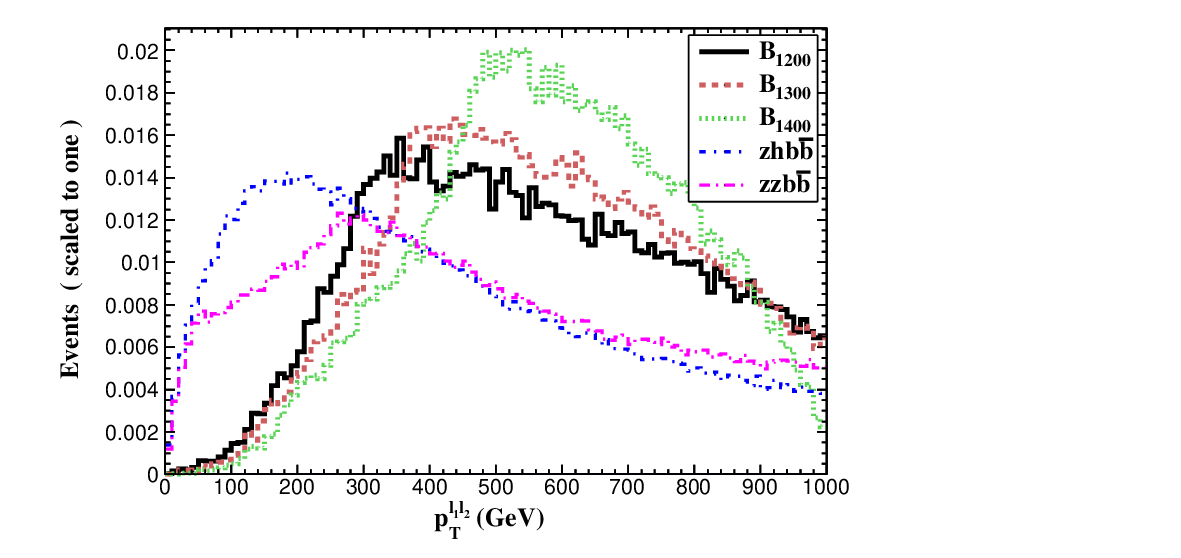}}
\centerline{\hspace{2.0cm}\epsfxsize=9cm\epsffile{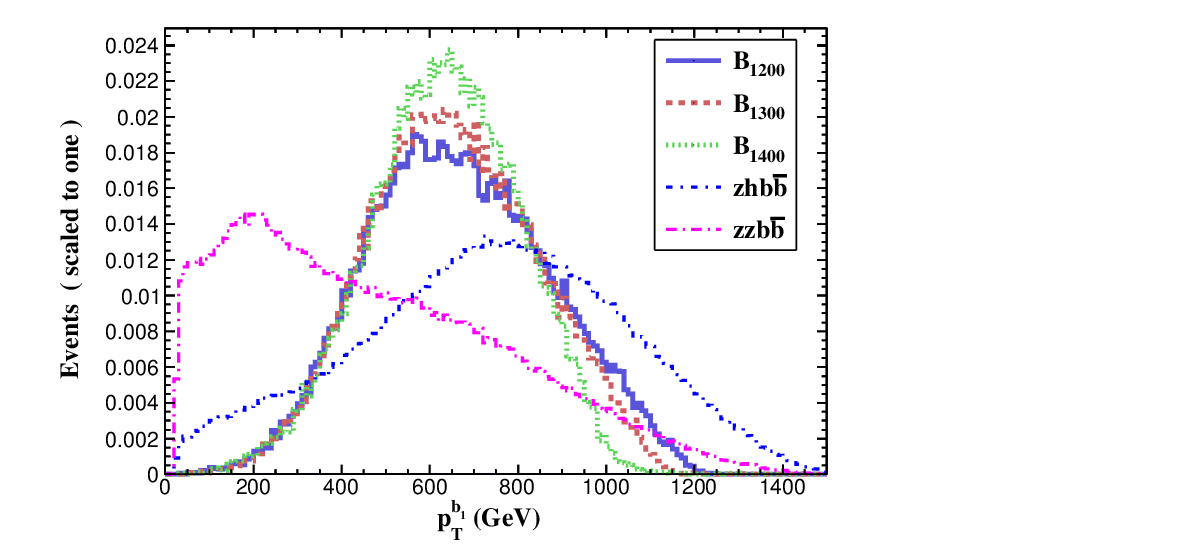}
\hspace{-2.0cm}\epsfxsize=9cm\epsffile{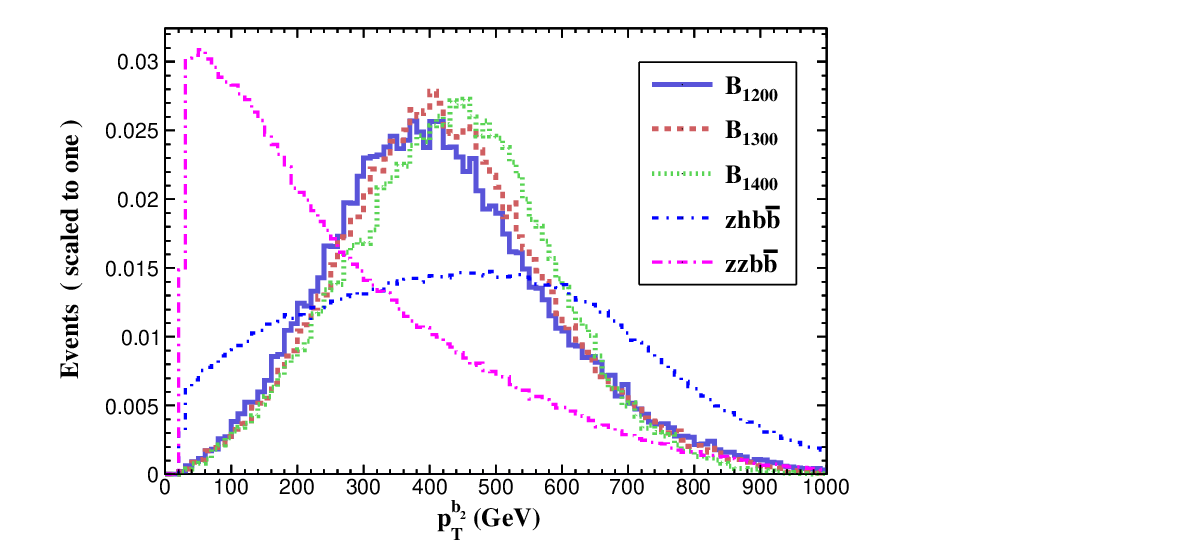}}
\centerline{\hspace{2.0cm}\epsfxsize=9cm\epsffile{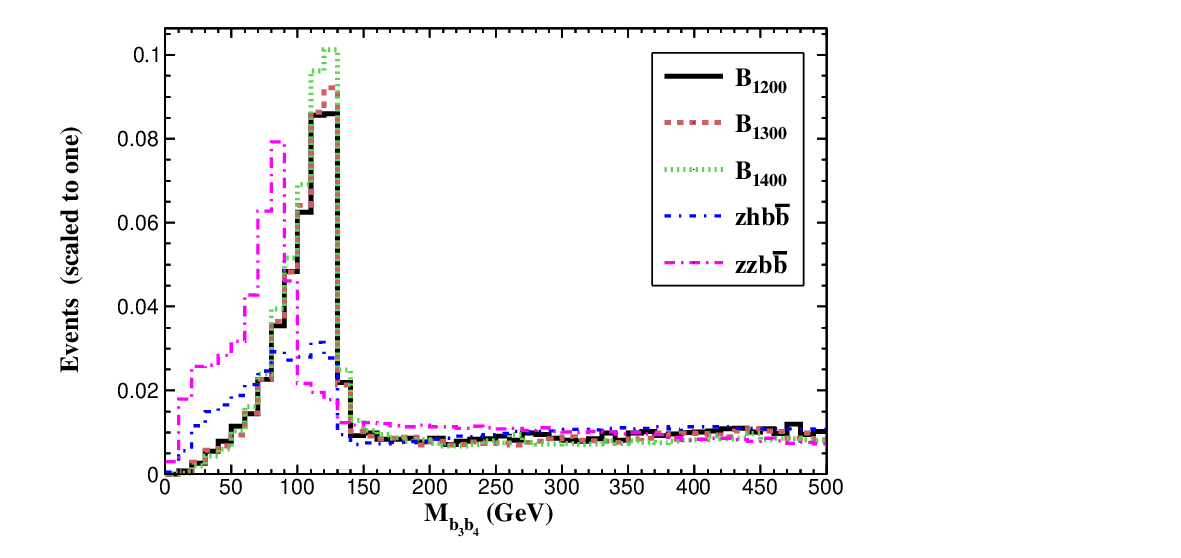}
\hspace{-2.0cm}\epsfxsize=9cm\epsffile{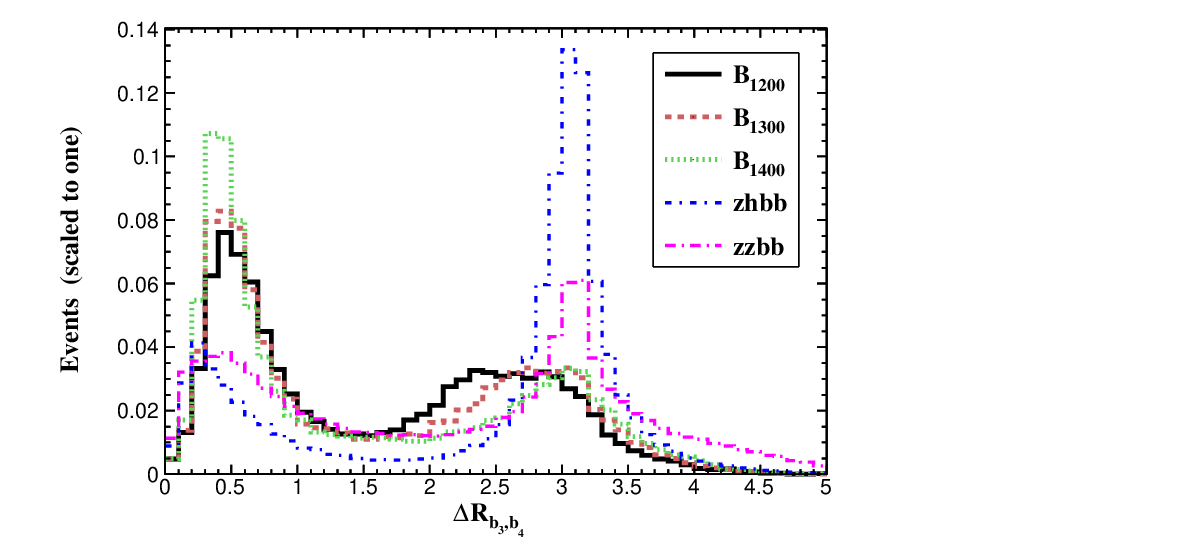}}
\caption{Normalized distributions for the signals (with $m_{B}$=1200, 1300, 1400 GeV) and SM backgrounds. }
\label{distribution1}
\end{center}
\end{figure*}

In Fig.~\ref{distribution1}, we plot  differential distributions  for  signals and SM backgrounds, including the invariant mass distribution for the $Z$ boson~($M_{\ell_{1}\ell_{2}}$), the transverse momentum distributions of the leading and sub-leading leptons~($p_{T}^{\ell_{1}\ell_{2}}$),  the transverse momentum distributions of the leading and sub-leading $b$-jets~($p_{T}^{b_{1}}, p_{T}^{b_{2}}$),  the invariant mass distribution for the $Z$ boson or Higgs boson ($M_{b_{3}b_{4}}$), and the separations $\Delta R_{b_{3},b_{4}}$.  For the signal, the leptons $\ell_{1}$ and $\ell_{2}$ are two OSSF leptons that are assumed to be the product of the $Z$ boson decay, and at least two $b$-tagged jets are assumed to be the product of one $Z$ boson or a Higgs boson decay.
Based on these kinematic distributions, we can impose
 the following set of cuts:
 \begin{itemize}
 \item Cut-1: There are exactly two isolated leptons~($N(\ell)=2$), and the invariant mass of the $Z$ boson is required to have $|M_{\ell_{1}\ell_{2}}-m_{Z}|< 10 \rm ~GeV$, and the transverse momenta of two leptons  are required $p_{T}^{\ell_{1}\ell_{2}}> 200 \rm ~GeV$.
\item Cut-2: There are exactly four $b$-tagged jets~($N(b)= 4$), and the transverse momenta of the leading and sub-leading $b$-jet are required $p_{T}^{b_{1}}> 400 \rm ~GeV$ and $p_{T}^{b_{2}}> 250 \rm ~GeV$.
\item Cut-3: The invariant masses of the $Z$ boson or the Higgs boson are required to have $50 \rm~GeV <M_{b_{3}b_{4}}< 150 \rm ~GeV$ with $\Delta R_{b_{3},b_{4}}< 1$.
\end{itemize}

\begin{table}[htb]
\centering %
\caption{Cut flow of the cross sections (in fb) for the signals with three typical VLQ-$B$ quark masses  and SM backgrounds.  \label{cutflow1}}
\vspace{0.2cm}
\begin{tabular}{p{1.6cm}<{\centering} p{1.7cm}<{\centering}  p{2.0cm}<{\centering} p{2.0cm}<{\centering}p{0.3cm}<{\centering}  p{1.5cm}<{\centering} p{1.7cm}<{\centering}}
\toprule[1.5pt]
 \multirow{2}{*}{Cuts}& \multicolumn{3}{c}{Signals}&&\multicolumn{2}{c}{Backgrounds}  \\ \cline{2-4}  \cline{6-7}
&1200 GeV & 1300 GeV & 1400 GeV  && $Zhb\bar{b}$ &$ZZb\bar{b}$\\    \cline{1-7} \midrule[1pt]
Basic&0.031&0.024&0.015&&0.005&0.004\\
Cut-1&0.023&0.018&0.012&&0.0025&0.0021\\
Cut-2&0.012&0.009&0.006&&0.0012&0.00055\\
Cut-3&0.0043&0.0036&0.0026&&0.00026&0.00015\\
\bottomrule[1.5pt]
\end{tabular}
 \end{table}

We present the cross sections of three typical signals ($m_B=1200, 1300, 1400$ GeV) and the relevant
backgrounds after imposing
the cuts in Table~\ref{cutflow1}.
One can see that all the SM backgrounds are suppressed very efficiently, while the signals still have  relatively good efficiency at the end of the cut flow. The cross section of the  total SM background is  about $0.4\times 10^{-3}$~fb.

\subsection{The decay channel $Z\to \nu\bar{\nu}$ }
In this subsection, we analyze the signal and background events through the decay channel of the invisible decays $Z\to \nu\bar{\nu}$:
\begin{equation}
    e^{+}e^{-} \to B\bar{B} \to \left\{ \begin{array}{l} \left(b Z\right)\,\left(\bar{b}Z\right) \\ \left(b Z\right)\,\left(\bar{b}h\right) \\ \left(b h\right)\,\left(\bar{b}Z\right) \end{array}\right. \to 4b+\slashed E_{T}.
\end{equation}
%%%%%%%%%%%%%%%%%%%%%%%%%%%%%%%%%%%%%%%%%%%%%%%%%%
\begin{figure*}[b]
\begin{center}
\centerline{\hspace{2.0cm}\epsfxsize=9cm\epsffile{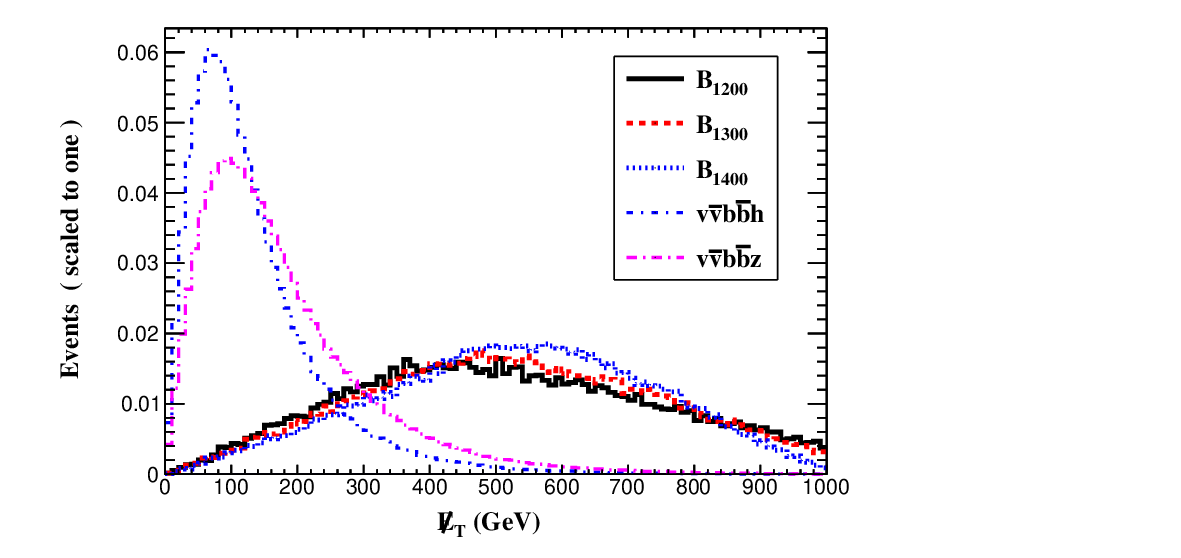}
\hspace{-2.0cm}\epsfxsize=9cm\epsffile{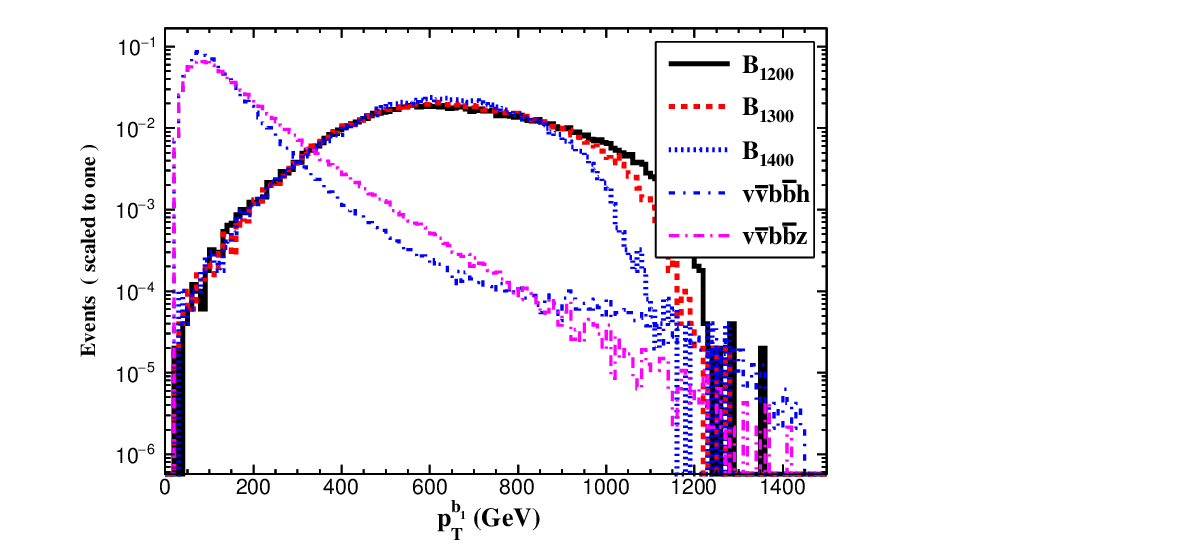}}
\centerline{\hspace{2.0cm}\epsfxsize=9cm\epsffile{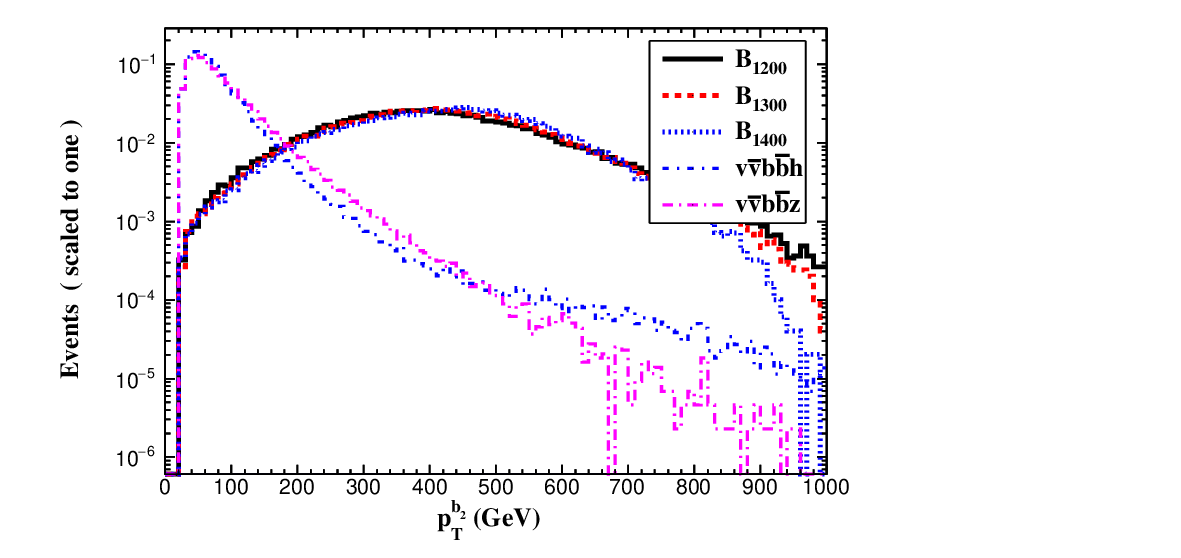}
\hspace{-2.0cm}\epsfxsize=9cm\epsffile{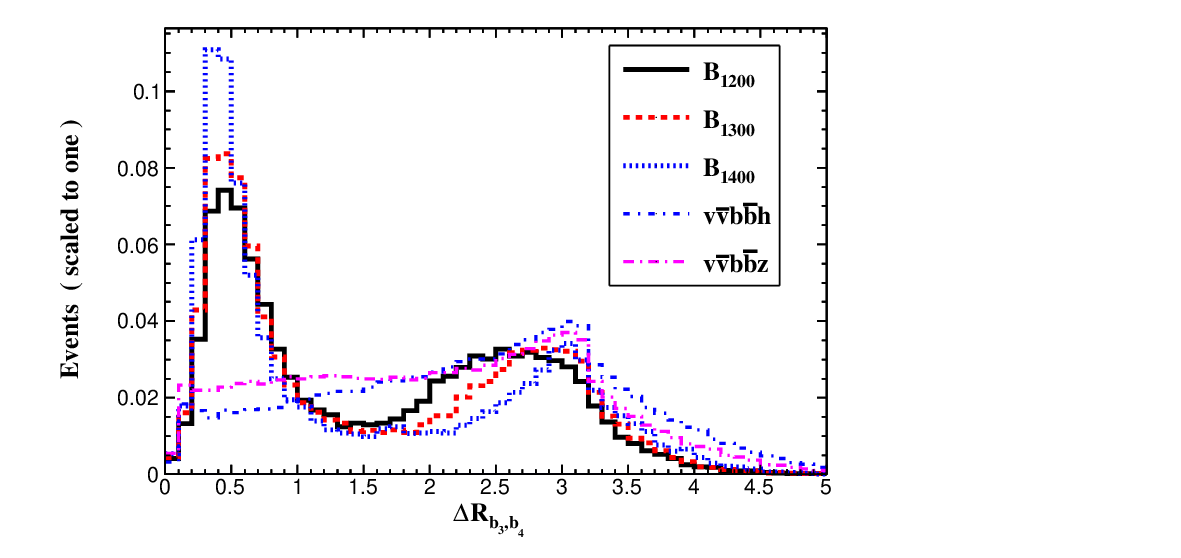}}
\centerline{\hspace{2.0cm}\epsfxsize=9cm\epsffile{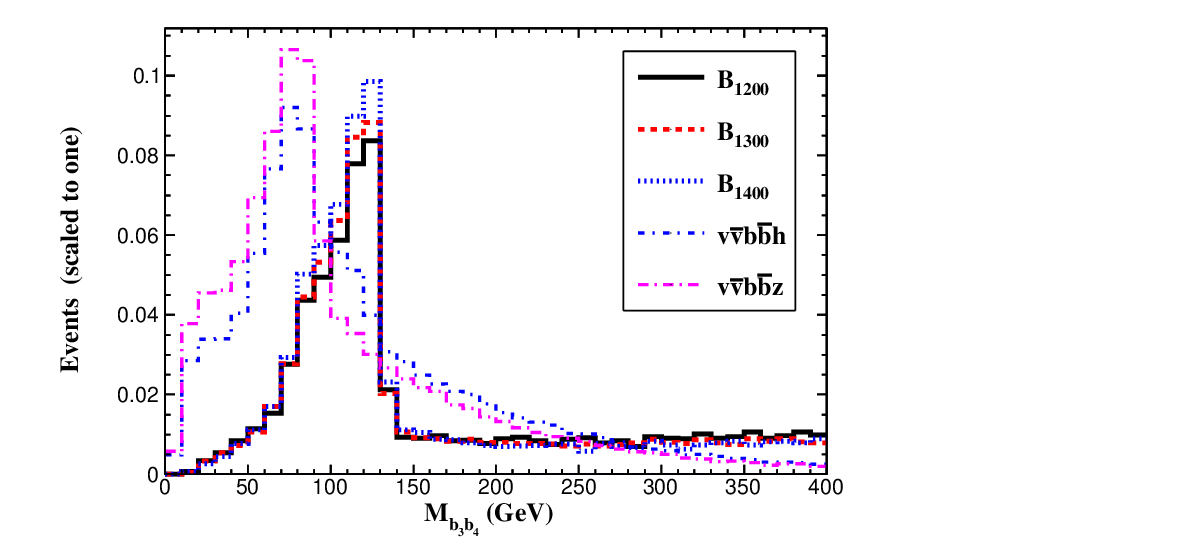}
\hspace{-2.0cm}\epsfxsize=9cm\epsffile{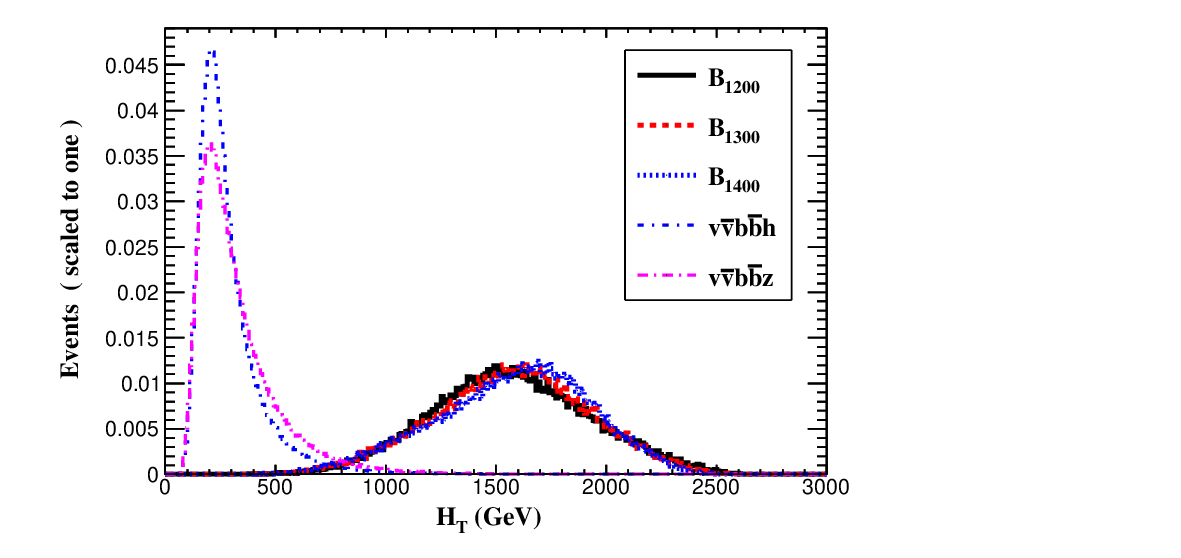}}
\caption{Normalized distributions for the signals (with $m_{B}$=1200, 1300, 1400 GeV) and SM backgrounds. }
\label{distribution2}
\end{center}
\end{figure*}

For this channel, the main SM backgrounds come from the processes
$e^{+}e^{-}\to \nu\bar{\nu}b\bar{b}Z$ and $e^{+}e^{-}\to \nu\bar{\nu}b\bar{b}h$ with the cross sections of 21.7~fb and 2.5~fb, respectively. Note that the contributions from the processes $e^{+}e^{-}\to Zhb\bar{b}$, $e^{+}e^{-}\to ZZb\bar{b}$, $e^{+}e^{-}\to Zhh$, $e^{+}e^{-}\to ZZh$, $e^{+}e^{-}\to ZZZ$, $e^{+}e^{-}\to \nu_{e}\bar{\nu}_{e}hh$, $e^{+}e^{-}\to \nu_{e}\bar{\nu}_{e}Zh$, and $e^{+}e^{-}\to \nu_{e}\bar{\nu}_{e}ZZ$ are also included with the decay modes $Z\to \nu\bar{\nu}$, $Z\to b\bar{b}$, and $h\to b\bar{b}$.

Obviously, the signal events should contain large missing transverse energy $\slashed E_{T}$  from the boosted $Z$ boson. Furthermore, at least two $b$-tagged jets are coming from the $Z$ boson or Higgs boson decay.
In order to obtain some hint of further cuts for reducing the SM backgrounds, we analyzed the normalized distributions of the missing transverse energy $\slashed E_{T}$, the transverse momentum distributions of the leading and sub-leading $b$-jets $p_{T}^{b_{1},b_{2}}$, the separations $\Delta R_{b_{3},b_{4}}$, the invariant mass distribution $M_{b_{3}b_{4}}$, and the scalar sum of the transverse energy of all final-state jets $H_{T}$ for  signals and SM backgrounds as shown in Fig.~\ref{distribution2}.
Based on these kinematic distributions, a set of further cuts are given as:
 \begin{itemize}
\item Cut-1: The transverse missing energy is required $\slashed E_{T}> 200 \rm ~GeV$.
\item Cut-2: Any electrons and muons are forbidden~($N(\ell) =0$) and there are exactly four $b$-tagged jets~($N(b) =4$). Furthermore, the transverse momenta of the leading and sub-leading $b$-jets are required $p_{T}^{b_{1}}> 300 \rm ~GeV$ and $p_{T}^{b_{2}}> 150 \rm ~GeV$. The invariant masses of the remaining two $b$-tagged jets are required to have $50 \rm~GeV <M_{b_{3}b_{4}}< 150 \rm ~GeV$ with $\Delta R_{b_{3},b_{4}}< 1$.
\item
Cut-3: The scalar sum of the transverse energy of all final-state jets $H_{T}> 1000 \rm ~GeV$.
\end{itemize}

\begin{table}[htb]
\centering %
\caption{Cut flow of the cross sections (in fb) for the signals with three typical VLQ-$B$ quark masses and SM backgrounds.  \label{cutflow2}}
\vspace{0.2cm}
\begin{tabular}{p{1.6cm}<{\centering} p{1.7cm}<{\centering}  p{2.0cm}<{\centering} p{2.0cm}<{\centering}p{0.3cm}<{\centering}  p{1.5cm}<{\centering} p{1.7cm}<{\centering}}
\toprule[1.5pt]
 \multirow{2}{*}{Cuts}& \multicolumn{3}{c}{Signals}&&\multicolumn{2}{c}{Backgrounds}  \\ \cline{2-4}  \cline{6-7}
&1200 GeV & 1300 GeV & 1400 GeV  && $\nu\bar{\nu}b\bar{b}h$ &$\nu\bar{\nu}b\bar{b}Z$\\    \cline{1-7} \midrule[1pt]
Basic&0.093&0.071&0.044&&2.44&4.51\\
Cut-1&0.085&0.066&0.042&&0.44&1.36\\
Cut-2&0.016&0.013&0.0093&&0.0044&0.011\\
Cut-3&0.015&0.013&0.0088&&0.0019&0.0037\\
\bottomrule[1.5pt]
\end{tabular}
 \end{table}

We summarize the cross sections of three typical signals ($m_B=1200, 1300, 1400$ GeV) and the relevant
backgrounds after imposing
the cuts in Table~\ref{cutflow2}.
One can see that the total SM backgrounds are suppressed very efficiently, with a cross section of about $5.6\times 10^{-3}$~fb.

\subsection{Discovery and exclusion significance }

In order to analyze the observability,
we use the median significance  to estimate the expected discovery and exclusion significance~\cite{Cowan:2010js}:
\be
\mathcal{Z}_\text{disc} &=
  \sqrt{2\left[(s+b)\ln\left(\frac{(s+b)(1+\delta^2 b)}{b+\delta^2 b(s+b)}\right) -
  \frac{1}{\delta^2 }\ln\left(1+\delta^2\frac{s}{1+\delta^2 b}\right)\right]} \\
   \mathcal{Z}_\text{excl} &=\sqrt{2\left[s-b\ln\left(\frac{b+s+x}{2b}\right)
  - \frac{1}{\delta^2 }\ln\left(\frac{b-s+x}{2b}\right)\right] -
  \left(b+s-x\right)\left(1+\frac{1}{\delta^2 b}\right)},
 \ee
where $ x=\sqrt{(s+b)^2- 4s \delta^2 b^2/(1+\delta^2 b)}$,
 $s$ and $b$ are the numbers of signal and background events at a given luminosity, respectively, and $\delta$ is the percentage systematic error on the SM
background estimate.
In the limit of
$\delta \to 0$,  these expressions  can be simplified as
\be
 \mathcal{Z}_\text{disc} &= \sqrt{2[(s+b)\ln(1+s/b)-s]}, \\
 \mathcal{Z}_\text{excl} &= \sqrt{2[s-b\ln(1+s/b)]}.
\ee
It is instructive to acknowledge systematic uncertainties at the experiment
which can effect our results.  To show this, we include a 10\% systematic uncertainty alongside the null systematic uncertainty results.

%%% Fig.5 %%%%%%%%%%%%%%%%%%%%
\begin{figure}[htb]
\begin{center}
\vspace{-0.5cm}
\centerline{\epsfxsize=8cm \epsffile{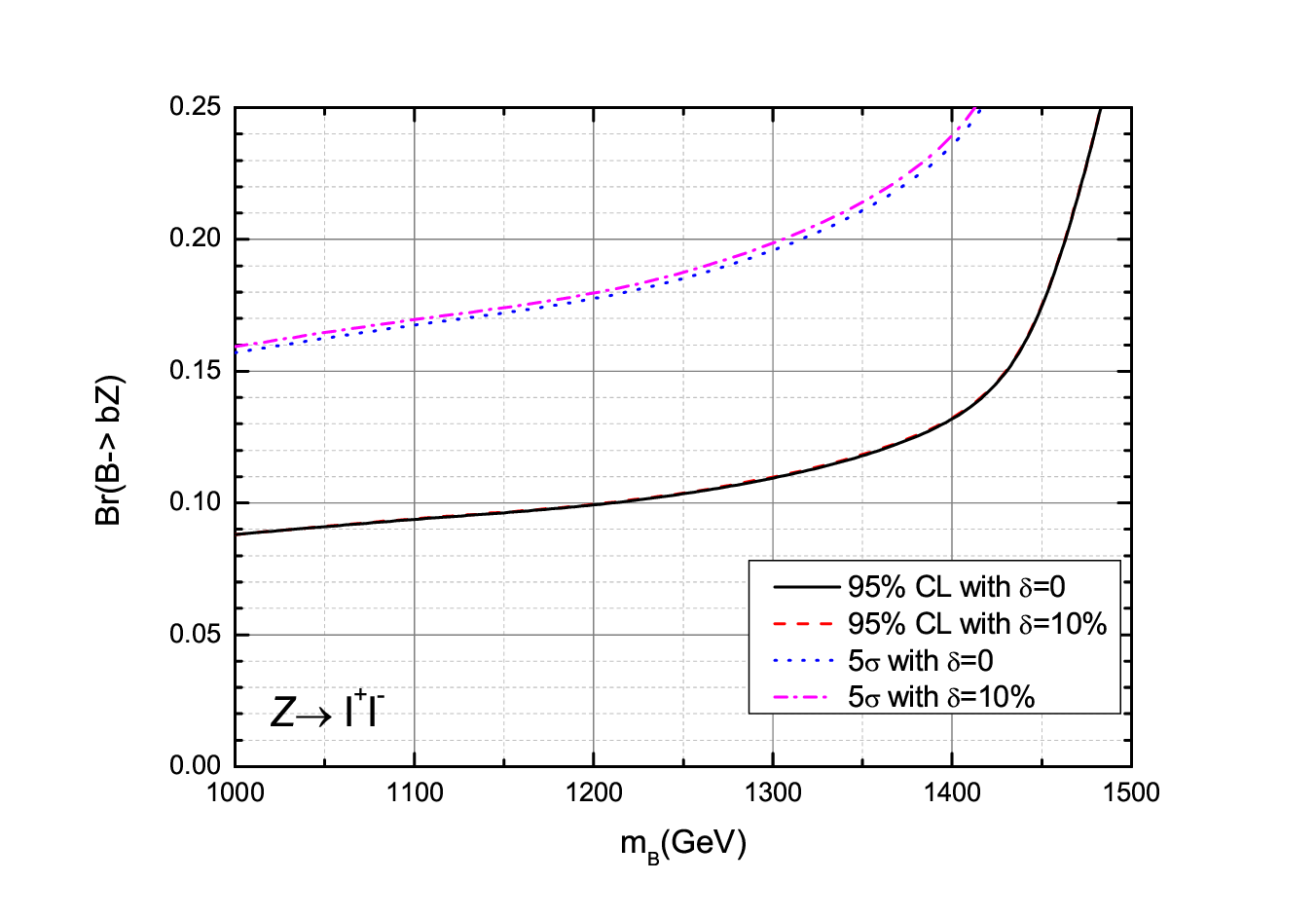}\epsfxsize=8cm \epsffile{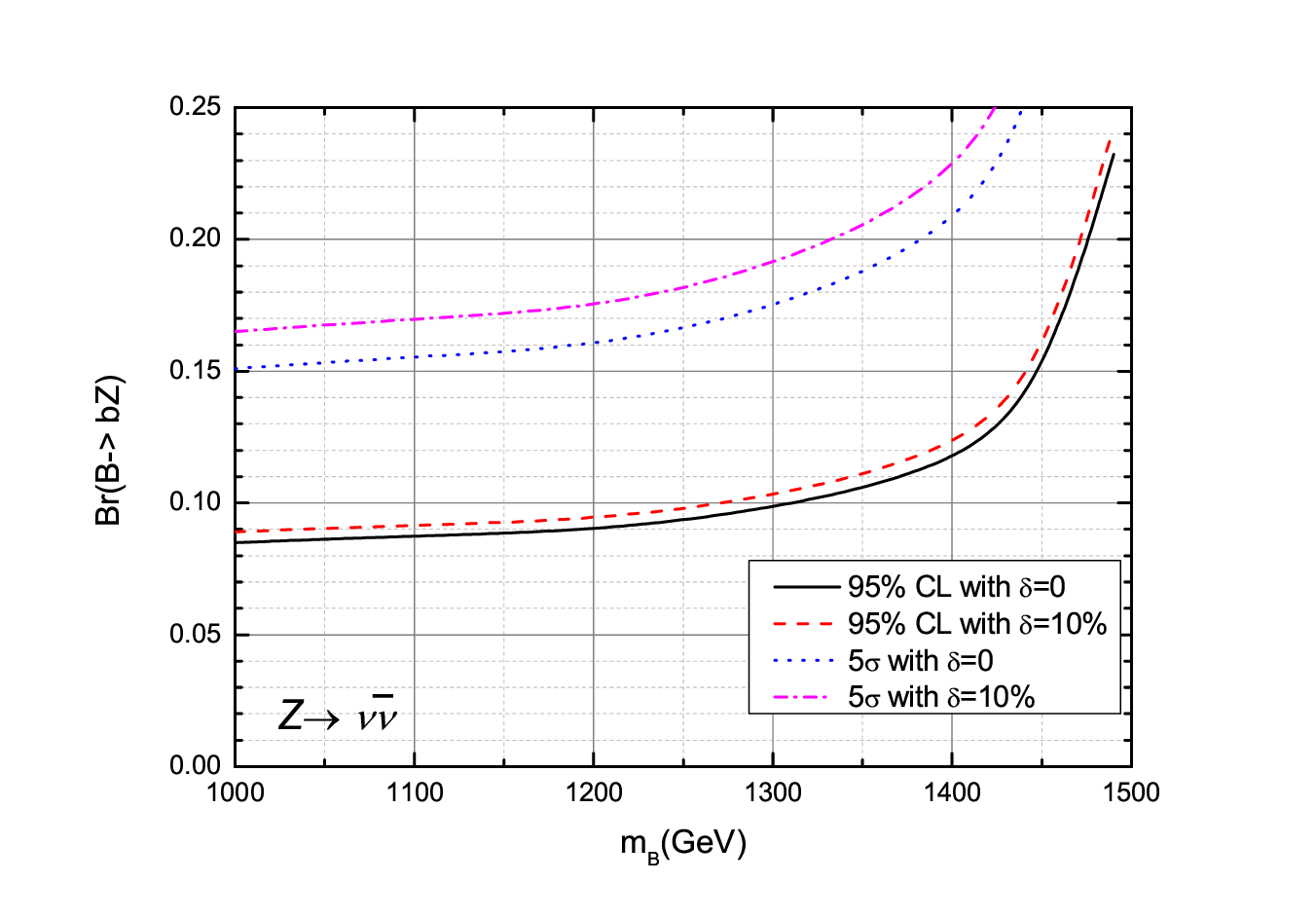}}
\caption{Exclusion limit~(at $95\%$ CL) and discovery prospects  (at $5\sigma$) contour plots for the $Z\to \ell^{+}\ell^{-}$ decay channel~(left), and for $Z\to \nu\bar{\nu}$ decay channel~(right) in the ${\rm Br}(B\to bZ)-m_{B}$ planes at the future 3-TeV CLIC with an integrated luminosity of 5 ab$^{-1}$.  }
\label{fig-ss}
\end{center}
\end{figure}

In Fig.~\ref{fig-ss}, we plot the 95\% CL exclusion limit and $5\sigma$ sensitivity reaches for ${\rm Br}(B\to bZ)$ as a function of $m_B$ at the 3-TeV CLIC with an integrated luminosity of 5 ab$^{-1}$ for two decay channels with the aforementioned two systematic error cases of  $\delta=0$ and $\delta=10\%$. We find that with a realistic 10\% systematic error, the sensitivities are slightly weaker than those without any systematic error. For the $Z\to \ell^{+}\ell^{-}$ decay channel,
the VLQ-$B$ quark can be excluded in the region of ${\rm Br}(B\to bZ)\in [0.09, 0.27]$ and $m_B\in$ [1000, 1490]~GeV  at the 3-TeV CLIC with an integrated luminosity of 5 ab$^{-1}$, while the discover region can reach  ${\rm Br}(B\to bZ)\in [0.16, 0.25]$ and $m_B\in$ [1000, 1420]~GeV.
For the $Z\to \nu\bar{\nu}$ decay channel, the VLQ-$B$ quark can be excluded in the region of ${\rm Br}(B\to bZ)\in [0.09, 0.24]$ and $m_B\in$ [1000, 1490]~GeV, and the discover region can reach  ${\rm Br}(B\to bZ)\in [0.165, 0.25]$ and $m_B\in$ [1000, 1430]~GeV  at the 3-TeV CLIC with an integrated luminosity of 5 ab$^{-1}$.

%%% Fig.7 %%%%%%%%%%%%%%%%%%%%
\begin{figure}[htb]
\begin{center}
\vspace{-0.5cm}
\centerline{\epsfxsize=9cm \epsffile{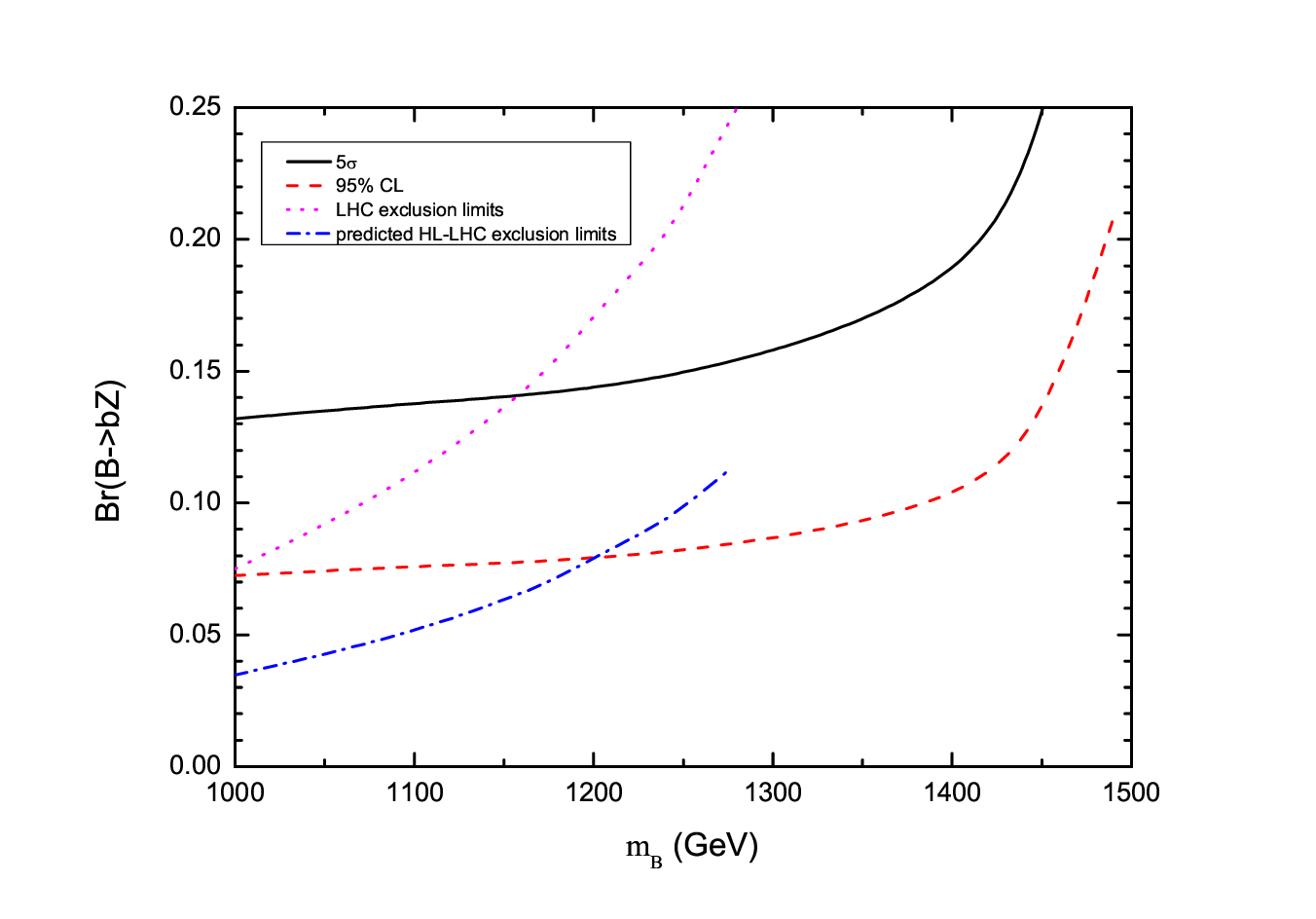}}
\caption{Combined  exclusion limit (at 95\% CL)  and discovery prospects (at $5\sigma$) contour plots for the signal in ${\rm Br}(B\to bZ)-m_{B}$ planes at 3~TeV CLIC with an integral luminosity of 5 ab$^{-1}$ with $\delta=10\%$. }
\label{fig7}
\end{center}
\end{figure}

Next, we combine the significance with
$\mathcal{Z}_\text{comb}=\sqrt{\mathcal{Z}^{2}_{\ell\bar{\ell}}+\mathcal{Z}^{2}_{\nu\bar{\nu}} }$ by using the results from the above two decay channels  with $\delta=10\%$. One can see that the singlet VLQ-$B$ can be excluded in the regions of ${\rm Br}(B\to bZ)\in [0.073, 0.21]$ and $m_B\in$ [1000, 1490]~GeV at the 3-TeV CLIC with integrated luminosity of 5 ab$^{-1}$, while the discover regions can reach ${\rm Br}(B\to bZ)\in [0.13, 0.23]$ and $m_B\in$ [1000, 1450]~GeV. For comparison,
we also present the observed 95\% CL exclusion limits at the 13-TeV LHC  and the predicted  exclusions at the future HL-LHC with an integrated luminosity of 3000~fb$^{-1}$. For simplicity, the exclusion limits for the HL-LHC are obtained by scaling the current LHC limits with the increased luminosity. We can observe that the future CLIC with $\sqrt{s}=3 \rm ~TeV$ and an integrated luminosity of 5~ab$^{-1}$ could provide better sensitivity than that reported in current experimental searches, and even better sensitivity than the future  HL-LHC in some mass regions of VLQ-B (e.g., 1200-1500 GeV).

\section{Conclusion}
In this work, we have concentrated on the pair production of the singlet VLQ-$B$ at the future 3 TeV CLIC in a simplified model.
With the increasing
 branching in the extra decay mode, the existing
limits on VLQs can be relaxed, so we first reinterpret  the latest mass-exclusion limits for VLQ-$B$ in terms of ${\rm Br}(B\to bZ)$.
Then we perform a full simulation for the signals and the relevant SM backgrounds on the final states including one $Z$ boson and four $b$-jets via two types of modes: $Z\to \ell^{+}\ell^{-}$ and $Z\to \nu\bar{\nu}$. We present the 95\% CL exclusion limits and $5\sigma$ discovery prospects in the parameter plane of the two variables ${\rm Br}(B\to bZ)$ and the  VLQ-$B$ masses  at the future 3-TeV CLIC with an integrated luminosity of 5 ab$^{-1}$.
The results show that the VLQ-$B$ quark can be excluded in the region of ${\rm Br}(B\to bZ)\in [0.073, 0.21]$ and $m_B\in$ [1000, 1490]~GeV  at the 3- TeV CLIC with integrated luminosity of 5 ab$^{-1}$, while the discover region can reach  ${\rm Br}(B\to bZ)\in [0.13, 0.23]$ and $m_B\in$ [1000, 1450]~GeV. We therefore expect that the signatures studied here will provide complementary information for detecting such VLQ-$B$, including non-standard decay modes, at the future 3-TeV CLIC.

\begin{acknowledgments}
The work is supported by the Project of Innovation and Entrepreneurship Training for College Students in Henan Province (202310478033).
\end{acknowledgments}

\end{document}